\documentclass{aa}

\usepackage{txfonts}
\usepackage{epsf}
\usepackage{graphicx}
\epsfclipon

\begin{document}
\title{Atmospheric dynamics in carbon-rich Miras}
\subtitle{I. Model atmospheres and synthetic line profiles}

\author{W.~Nowotny\inst{1} 
        \and B.~Aringer\inst{1}
        \and S.~H\"ofner\inst{2}
        \and R.~Gautschy-Loidl\inst{3}
        \and W.~Windsteig\inst{1}  }
\institute{Institut f\"ur Astronomie der Universit\"at Wien,
           T\"urkenschanzstra{\ss}e 17, A-1180 Wien, Austria
      \and Departement of Astronomy and Space Physics, Uppsala University, 
           Box 515, SE-75120 Uppsala, Sweden
      \and Froburgstrasse 43, CH-4052 Basel, Switzerland}
\offprints{W. Nowotny, \\ nowotny@astro.univie.ac.at}
\date{Received / Accepted }
\titlerunning{Atmospheric Dynamics in Carbon-rich Miras I.} 
\authorrunning{Nowotny et al.}

\abstract{
Atmospheres of evolved AGB stars are heavily affected by pulsation, dust formation and mass loss, and they can become very extended. Time series of observed high-resolution spectra proved to be a useful tool to study atmospheric dynamics throughout the outer layers of these pulsating red giants. Originating at various depths, different molecular spectral lines observed in the near-infrared can be used to probe gas velocities there for different phases during the lightcycle. Dynamic model atmospheres are needed to represent the complicated structures of Mira variables properly. An important aspect which should be reproduced by the models is the variation of line profiles due to the influence of gas velocities. Based on a dynamic model, synthetic spectra (containing CO and CN lines) were calculated, using an LTE radiative transfer code that includes velocity effects. It is shown that profiles of lines that sample different depths qualitatively reproduce the behaviour expected from observations.

   \keywords{Stars: late-type --
             Stars: AGB and post-AGB --
             Stars: atmospheres --
             Stars: carbon --
             Infrared: stars --
             Line: profiles   }
         }
\maketitle


\section{Introduction}      \label{s:intro}

Stars of low to intermediate masses at a late stage of their evolution are characterised by high luminosities, large extensions, low effective temperatures and pronounced variability. After core helium burning has ceased, they reach the evolutionary stage of the Asymptotic Giant Branch (AGB). Towards the end of this stage, repeated and explosive helium burning in a shell sets in (thermal pulses, TP). Nucleo-synthesis products, especially carbon but also s-process elements, can be mixed up from the stellar interior by deep-reaching convection (dredge-up) and change the atmospheric chemical composition (e.g. Busso et al. \cite{BusGW99}). Starting with an oxygen-rich (C/O$<$1) atmosphere at the onset of the AGB-evolution, the stars can become carbon-rich (C/O$>$1) during the TP-AGB phase. This results in a metamorphosis of the spectral appearance: the spectral type changes from M to C. 

Red giants become unstable to radial pulsation with often large amplitudes and long periods of a few 10 to several 100 days. Among these so-called Long Period Variables (LPV), Mira stars show a well-pronounced and regular variability. While their amplitudes reach several magnitudes in the visual, they decrease toward the infrared and bolometric light changes are of the order of 1$^{\rm mag}$.

Pulsations of the stellar interior have a strong effect on Mira atmospheres as they trigger shock fronts, which propagate outwards. This causes a levitation of the atmosphere, which consequently becomes very extended. The occurrence of shock fronts has been suspected from observations of emission lines around phases of maximum light (e.g. Richter et al. \cite{RichW01}, \cite{RWWBS03}) and by the detection of line doubling found for molecular lines in the near infrared (e.g. Hinkle et al. \cite{HinHR82}). In the cool, dense environments of red giant atmospheres, molecules can form efficiently, resulting in the typical line-rich late-type spectra. In the post-shock regions, dust grains can condense from these molecules (e.g. Sedlmayr \cite{Sedlm94}, Millar \cite{Milla04}). Like the molecular species, dust condensates also depend on the chemical composition of the atmosphere. For oxygen-rich objects, silicates (amorphous as well as crystalline) and refractory oxides (e.g. Mg-Al-oxides) are found, while for carbon-rich ones amorphous carbon and SiC can be observed (Molster \& Waters \cite{MolsW03}). Radiation pressure on dust particles is believed to be a major reason for the slow stellar wind from AGB stars leading to high mass loss rates (dust-driven winds; see e.g. Simis \& Wotike \cite{SimiW04}). A recent general overview of the complex phenomena occuring in AGB atmospheres was given by Gustafsson \& H\"ofner (\cite{GustH04} and references therein).

High-resolution spectroscopy in the near infrared (NIR) proved to be a very powerful tool to study dynamic processes throughout the atmospheres of these stars (Hinkle et al. \cite{HinHR82}), as discussed in detail in Sect.\,\ref{s:lineprofile}. The aim of this paper is to reproduce the diverse behaviours of several molecular lines by modelling synthetic line profiles from dynamic model atmospheres.


\section{Observed line profile variations in AGB stars} \label{s:lineprofile}

\subsection{Radial velocities from spectral features} \label{s:RVfeature} 

Spectral energy distributions of AGB stars peak in the NIR region around 1--3$\mu$m. The spectra are densely populated by numerous atomic and molecular absorption lines, requiring high resolution for detailed spectroscopic studies. Radial velocities (RV) derived from shifts in wavelength of spectral lines can provide clues to the velocites in the layers where these lines originate.

Line shifts in Mira spectra have been explored observationally for a long time. The first investigations of these variations were made in the blue and the visual spectral range (e.g. Joy \cite{Joy54}, Wallerstein \cite{Walle75}) revealing puzzling results, as outlined in Hinkle (\cite{Hinkl78}) and Hinkle et~al. (\cite{HinHR82}). Radial velocities from emission lines, observed around maximum brightness, were consistent with the picture of a pulsating photosphere  in which shockfronts propagate. However no large amplitude RV variations over the light cycle for absorption lines were found, as would be expected if the spectral variability is coupled to radial pulsations. Only a systematic red-shift to the center-of-mass radial velocity (CMRV) for almost all phases was derived (e.g. Fig.\,4 in Hinkle \& Barnes \cite{HinkB79b}). As these expected periodic RV variations were then found for other atomic and molecular absorption lines in the red spectral region (Sanford \cite{Sanfo50}) and later in the NIR (see below), it was suspected that sufficient optical depth in the visual prevents the blue-shifted components of absorption lines from being observable (Hinkle \& Barnes \cite{HinkB79b}, Hinkle \& Barnbaum \cite{HinkB96}).

Major advances in the studies of LPV atmospheric kinematics resulted from the availability of Fourier transform spectrometers. By applying high-resolution spectroscopy in the NIR, new insights could be gained. This spectral region is especially well suited for Miras, not only because their fluxes peak at these wavelengths but also because variability is less pronounced there compared to the visual (hence spectroscopic monitoring over the light cycle is easier, even during minimum phases).

The oxygen-rich Mira R\,Leo was first studied in detail using spectra in the range of 1.6--2.5$\mu$m for different phases. The results were presented in a series of papers by Hinkle (\cite{Hinkl78}) and Hinkle \& Barnes (\cite{HinkB79a}, \cite{HinkB79b}). They analyzed the behaviour of different molecular (CO, OH and H$_2$O) and atomic lines in this wavelength region and derived RVs. Hinkle et al. (\cite{HinHR82}, in the following HHR82) then presented a very thorough study for $\chi$\,Cyg, a Mira variable of spectral type S, with a time series of several spectra at 1.6--2.5$\mu$m (providing good coverage of the pulsational period) and a few at 4.6$\mu$m. From these observations it is known that different lines behave differently. Originating in various depths of the very extended AGB atmospheres, RVs derived from line shifts provide information about gas velocities in these layers for different phases. By combining results from divers spectral regions, Hinkle et al. were able to give an overall interpretation of the complex dynamics going on in the atmospheres of these stars during a lightcycle.

The variation of NIR lines was then explored further (larger samples of objects, different types of LPVs, different molecules, etc.) by several authors, e.g. Hinkle et al. (\cite{HinSH84}) or Lebzelter et al. (\cite{LebHH99}, \cite{LebHA01}). Wallerstein (\cite{Walle85}) showed that most of the layers sampled by NIR CO lines can be associated with RVs of selected lines in the optical region and attempted to derive a stratigraphy for $\chi$\,Cyg at different phases by combining optical and near infrared data. Recently, observations of spectral variations in the visual range have been  obtained by Alvarez et al. (\cite{AJPGF00}, \cite{AJPGF01a}, \cite{AJPGF01b}). From these they deduce a tomography of LPV atmospheres and present statistics on line-doubling among Miras.

\subsection{Molecular lines of CO in the NIR}   \label{s:COsection}

Although lines of several molecules and atoms were analysed, CO lines proved to be most useful for these kinds of investigations for several reasons: (i) both carbon and oxygen are abundant elements and CO has a large dissociation energy, (ii) CO molecules are formed and can be observed in all types of chemistry (M, S, C stars), (iii) being formed already in deep photospheric layers, CO is abundant and stable over the whole atmosphere and large parts of the CSE (eventually it is photo-dissociated by the interstellar radiation field), (iv) it experiences no depletion into dust, (v) three sequences of vibration-rotation bands with sufficient line strength are observable in atmospheric windows. The achieved results (especially from HHR82) demonstrate that different vibration-rotation bands of CO are ideal probes of various regions within extended Mira atmosphere. As HHR82 state, CO bands are useful, because the observable lines span a wide range of excitation energy and strength, and line lists of reasonable quality exist (line positions, oscillator strengths). Also, individual, unblended lines can be identified in the NIR, as well as points of the continuum.

CO $\Delta v$=3 absorption lines, found in the H-band at $\lambda$$\approx$1.6$\mu$m, show blue/red--shifts coupled to the photometric variability and even line-doubling around visual maximum. The typical \textit{S-shaped}, discontinuous RV curves (e.g. Fig.\,12 of HHR82) can be regarded as an indication of atmospheric motion driven by pulsation and accompanied by shockfronts running through the photosphere. This behaviour can also be found for high-excitation lines of CO $\Delta v$=2 in the K-band at $\lambda$$\approx$2.3$\mu$m, but the low-excitation lines there show smaller variations not necessarily related to the pulsation cycle. While the first originate in pulsating layers, the latter are formed further out in the dust-forming layers. CO $\Delta v$=1 lines observable in the M-band at $\lambda$$\approx$4.6$\mu$m  sample the outflow in the outer parts of the atmosphere and show roughly the same RV at any phase. Figure\,\ref{f:regions} gives an overview of this scenario.

In Fig.\,1 of Lebzelter \& Hinkle (\cite{LebzH02}) a composite of RV-curves derived from CO $\Delta v$=3 lines of all Miras observed so far was put together, showing a rather uniform picture (shape, amplitudes of about 20--30\,km\,s$^{-1}$ around the CMRV) independent of spectral type, period or metallicity. Apart from this common characteristic, line profiles can become highly asymmetric; their shapes and variations appear different in each observed star. This is even more pronounced around 4\,$\mu$m (Lebzelter et al. \cite{LebHA01}).

\subsection{Carbon-rich stars}  \label{s:cstars}

This paper is dedicated to the exploration of C-rich atmospheres, as the formation and evolution of C-rich dust is much better understood and thus implemented in the models used here (cf. Sect.\,\ref{s:modelling} and H\"ofner et al.\,\cite{HoGAJ03}). 

Most of the stars observed possess oxygen-rich atmospheres. Much less observational material exists for Miras of spectral type C. After the early publication by Sanford (\cite{Sanfo50}), carbon-rich AGB stars were later studied by Phillips \& Freedman (\cite{PhilF69}), Hirai (\cite{Hirai83}), Keady et al. (\cite{KeaHR88}), Barnbaum (\cite{Barnb92a}, \cite{Barnb92b}) and Barnbaum \& Hinkle (\cite{BarnH95}). S\,Cep is the only C-rich Mira for which extensive time series IR spectroscopy has been obtained so far. The analysis was presented by Hinkle \& Barnbaum (\cite{HinkB96}, in the following HB96). S\,Cep was therefore chosen as the reference object for our modelling. Spectral features of CN are prominent in visual and NIR spectra of C stars and can in addition to CO be used to infer velocity information.


\section{Modelling line profile variations} \label{s:lineprofilesynth}

If the behaviour of different molecular features can be reproduced by synthetic spectra, this would be an indicator of the quality of the models used and a confirmation of the correctness of the ideas about the dynamic processes going on in the outer layers of Miras. On the other hand, the interpretation of complex multi-component line profiles, caused by the non-monotonic velocity fields throughout Mira atmospheres, will be aided by such computations.

To model the line profiles discussed above, an atmospheric structure is needed  for which detailed radiative transfer including velocities can be solved to obtain synthetic spectra. Modelling the cool and very extended Mira atmospheres remains a challenging problem as they have complex, temporally variable structures. Hydrostatic model atmosphere codes were only able to reproduce observed spectra for AGB stars with \textit{mild} pulsations and higher effective temperatures, e.g. J{\o}rgensen et al. (\cite{JorHL00}) or Loidl~et~al. (\cite{LoiLJ01}) for C-rich stars and Aringer et al. (\cite{AriKL97}, \cite{AriKJ02}) for O-rich stars. However, this is not an adequate approach for red giants with more pronounced pulsation and mass loss. It is even impossible to fit observations for these objects at various phases by different hydrostatic models. Dynamic models are in particular required to reproduce phenomena such as Doppler-shifted spectral lines.

Hill \& Willson (\cite{HillW79}) created the first hydrodynamic atmospheric model to describe the observed velocity changes and emission features. Since then, more sophisticated models for the atmospheres of LPVs and dust--driven winds have been developed by different groups. Reviews on the different approaches were given by  H\"ofner (\cite{Hoefn99}) and Woitke (\cite{Woitk98}, \cite{Woitk03}). Considerable progress has been achieved in this field during the last few years through comparison with IR spectroscopic data (ISO). In general, dynamic models are now approaching quantitative agreement with observations (H\"ofner et al. \cite{HoGAJ03}, Gautschy-Loidl et al. \cite{GaHJH04}, H\"ofner et al. \cite{HGANH04}). In contrast, there have been only a few attempts to model line profiles and their variations for pulsating AGB atmospheres in the past, which shall be summarised in the following.

\subsection{Pulsating model atmospheres} \label{s:bessell} 

Based on simple synthetic atmospheric structures of Bessell et al. (\cite{BeBSW89a}), Bessell et al. (\cite{BesBS88}) presented NIR CO lines that show asymmetries and line doubling as in the observations. Using the more advanced dynamic models of Bessell \& Scholz (\cite{BessS89}), Scholz (\cite{Schol92}) showed how velocities in Mira photospheres distort line profiles of different species and affect measurements of equivalent widths and curves of growth for chemical analyses from these spectra. Starting with the same models, Bessell et al. (\cite{BesSW96}) for the first time calculated synthetic line profiles (CO $\Delta v$=2, Fe\,I) for different phases, which showed qualitative similarities with some observed features (line-shifts, doubling, velocity amplitudes). Scholz\,\&\,Wood (\cite{SchoW00}) then presented similar calculations for different lines (CO $\Delta v$=2,3, OH $\Delta v$=2) and derived conversion factors relating RVs derived from Doppler-shifted lines to actual gas velocities above and below the shock wave running through the atmosphere.

A unique feature of the dynamical models by Bessell et al. (\cite{BesSW96}) and Hofmann et al. (\cite{HofSW98}) is that they are based on self-excited pulsation models and should therefore be suitable to represent the inner pulsating layers of a Mira atmosphere. Bessell et al. (1996) fitted the temporal variations of sub-photospheric layers of the pulsation models with Fourier series and used these fits as inner boundary conditions for dynamical calculations of the outermost layers with higher spatial resolution. Hofmann et~al. (\cite{HofSW98}) directly used the structures from their new pulsation models but followed the same procedure as Bessell et al. for the older models. In both cases, the density and velocity structures of the dynamical models were used directly for the calculation of observable properties. Based on the densities, however, new values for the gas temperatures were derived using radiative transfer in about 70 frequencies before computing the detailed spectra. This was necessary because the temperatures resulting from the grey dynamical calculations were too unrealistic. A drawback of this procedure is that it neglects the feedback of non-grey effects on the density and dynamics which may be crucial. Furthermore, different sources of opacity data were used in the different steps which adds to possible inconsistencies. In contrast to that, the dynamical models used in this paper are based on a simultaneous solution of hydrodynamics and frequency-dependent radiative transfer leading to consistent dynamical density-temperature structures. It should also be pointed out that our models include a time-dependent description of dust condensation and allow for the formation of dust-driven stellar winds whereas the models of Bessell et al. (\cite{BesSW96}) and Hofmann et al. (\cite{HofSW98}) are pure atmospheric models (no dust and therefore no mass loss). Also the latter are focused on O-rich chemistry; no dynamic models for carbon stars have been computed by Bessell, Hofmann and collaborators so far.

\subsection{Dust-driven wind models} \label{s:berlin}

Improving the semi--empirical approach of Keady et al. (\cite{KeaHR88}) for spectral synthesis, Winters et al. (\cite{WiKGS00}) presented synthetic fundamental and first overtone CO line profiles and compared them with observed ones for the obscured, C-rich Mira IRC+10216. Calculating synthetic spectra for different phases of a set of dynamic models with different parameters, they demonstrated the influence of mass loss rates (and hence dust optical depth) on the shape of the profiles for observations at different pulsational periods. By suppressing CO absorption for different layers, Winters et al. investigated their contribution to the final profiles.

The models used by Winters et al. (\cite{WiKGS00}) are based on a detailed  description of dust formation and stellar wind dynamics and are well suited for describing stars with heavy mass loss and spectral features coming from the optically thick and dusty outflow. This allowed them to interpret variations observed in the first overtone of CO as the formation and dynamics of discrete dust shells predicted by the models. On the other hand, these models contain only a crude description of the stellar atmosphere, using grey radiative transfer and no molecular opacities. This leads to considerable differences in the density-temperature structures, in particular in deep photospheric layers and further out in the upper atmospheric layers where the dust formation starts. This may be one of the reasons behind the problems of fitting both the global spectral energy distribution and the line profiles with one consistent model, as mentioned by Winters et al. (\cite{WiKGS00}). They rescaled the density of the wind to get line profiles comparable with observations but used the original model to compute the spectral energy distribution. Due to the uncertatinties in mass loss rates derived from observations, no detailed fitting of the observed SED with the purpose of constraining the densities was performed. Compared to the models of Winters et al., the dynamical models  used here combine a more realistic non-grey description of the atmosphere with a similar treatment of dust formation and wind dynamics. This allows us to study also stars with moderate mass loss rates that are not completely obscured by dust, and lines which originate in very different geometrical depths in the atmosphere or wind.


\section{Combined atmosphere and wind models} \label{s:modelling}

The discussion in Sect.\,\ref{s:lineprofilesynth} shows that existing dynamical models that were used to study line profile variations in AGB stars fall into two groups: atmospheric models (e.g. Scholz \& Wood \cite{SchoW00}) or wind models (e.g. Winters et~al. \cite{WiKGS00}). The former deal with lines originating from the various layers of the pulsating atmosphere, while the lines studied with the latter models probe layers from the dust formation zone to the outflow region. The purpose of the new models presented in this paper is a consistent description of all these phenomena, i.e. simultaneous modelling of lines originating in various layers (from the deep photosphere out to the dust formation region and beyond to the stellar wind region), with one single dynamical model for a given star.\footnote{A common feature of all these dynamic models is that they are computed in spherical geometry; effects of asymmetry  therefore cannot be studied with these 1D-models.}

For the calculations presented here, dynamic model atmospheres as described in H\"ofner et al. (\cite{HoGAJ03}) were used. Starting with a hydrostatic initial model, the equations of hydrodynamics, frequency-dependent radiative transfer and time-dependent dust formation are solved simultaneously to get an adequate description of the highly dynamic AGB atmospheres. This results in a more realistic description of both the dust-free pulsating atmosphere and the dust-driven stellar wind compared to previous models. The pulsation is simulated by a variable inner boundary below the stellar photosphere (piston). Luminosity changes sinusoidally due to this varying boundary condition. See H{\"o}fner et al. (\cite{HoGAJ03}) for more details about the modelling method. 

\begin{figure}
   \resizebox{\hsize}{!}{\includegraphics{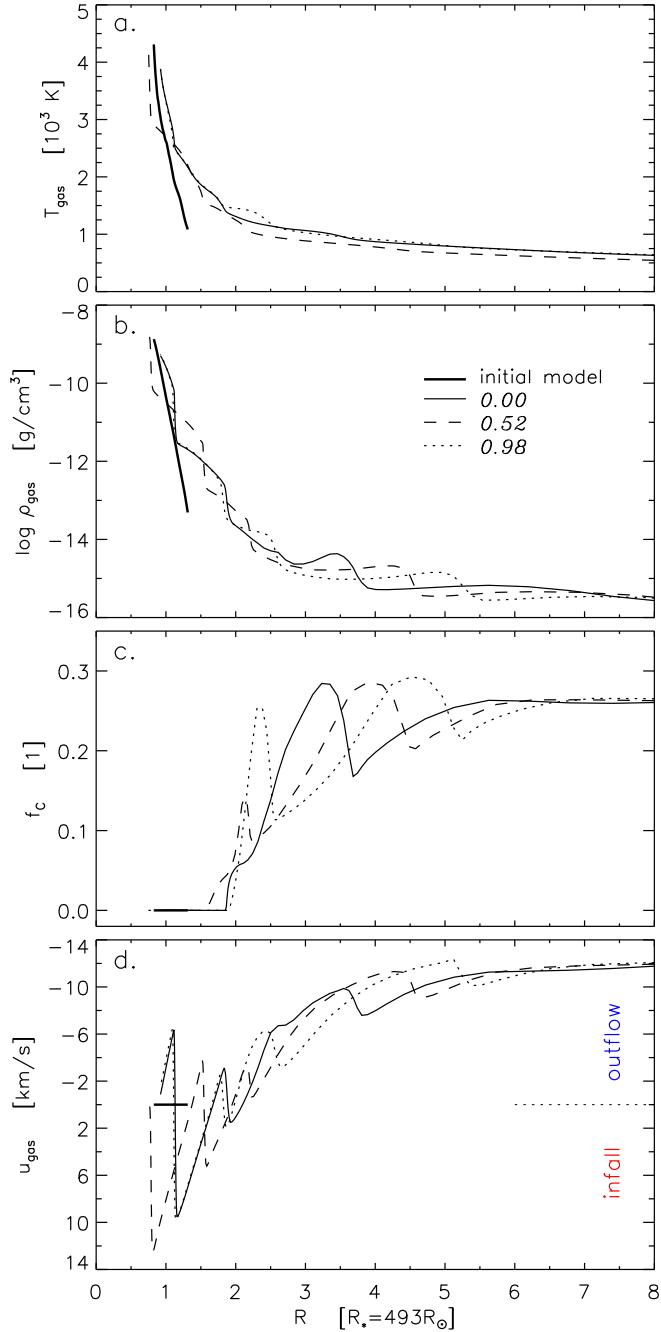}}
    \caption[]
    {Radial structures of the initial hydrostatic model and three 
    selected phases
    $\phi_{\rm bol}$ of the dynamic model atmosphere used here for calculating 
    synthetic spectra.
    Plotted are gas temperature (a), gas density (b), degree of condensation of
    carbon into dust (c) and gas velocity (d). 
    The convention of assigning negative velocities to outflowing matter was
    adopted from observational publications for compatibility 
    (cf. Sect.\,\ref{s:velocityconvention}).}
   \label{f:structure}
\end{figure}            

In contrast to O-rich objects, formation of polyatomic molecules and dust happens within similar temperature regimes for C-rich stars (Loidl et al. \cite{LoHJA99}). It is therefore necessary that opacities of molecules (not included in the models described in Sect.\,\ref{s:berlin}) and dust (not included in the models described in Sect.\,\ref{s:bessell}) are simultaneously treated in the computations of the model atmosphere to get realistic atmospheric structures and NIR spectra. In our current dynamical models, the formation, growth and evaporation of dust grains in the C-rich case is treated by the method described in Gauger et al. (\cite{GauGS90}). In the case of O-rich chemistry, on the other hand, dust is in our models at present simply included in the form of a parameterised opacity, due to uncertainties in connection with the mirco-physics of grain formation.\footnote{Self-consistent models for O-rich mass-losing Miras were presented by Jeong et al. (\cite{JeWLS03}).} Therefore we started our calculations with models for carbon-rich Miras.

\begin{figure}
   \resizebox{\hsize}{!}{\includegraphics{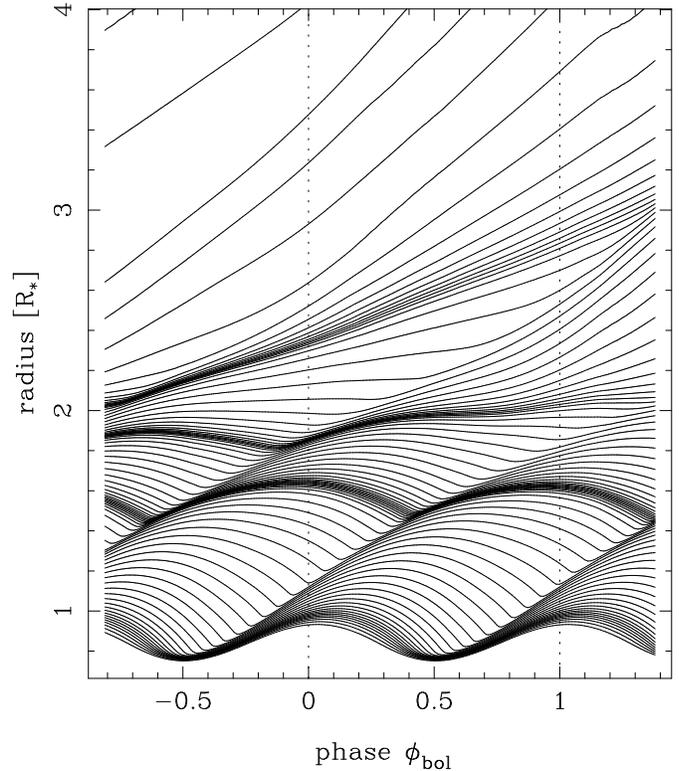}}
    \caption[]
    {Movement of mass shells with time at different depths of the dynamic model 
    atmosphere (cf. H\"ofner et al. \cite{HoGAJ03}, Fig.\,2). 
    Pulsation dominates the absolutely regular motion in the innermost parts 
    ($\approx$0.8--1.8\,R$_{\ast}$). Where infalling matter collides with
    outflowing gas from the next pulsation cycle, shockfronts emerge
    and propagate outwards.
    Starting at $\approx$2\,R$_{\ast}$, dust is formed in dense regions and
    radiation pressure on dust particles triggers the stellar wind. In these
    regions velocities are no longer periodic.
    From $\approx$4\,R$_{\ast}$ outwards the dust-driven wind dominates and 
    the model
    shows a stationary outflow.}
   \label{f:massenschalen}
\end{figure}

These models have proven already to represent observed properties of AGB stars rather well, such as the global shape of spectral energy distributions (H\"ofner et al. \cite{HoGAJ03}) and low--resolution spectra over a wide wavelength range (Gautschy-Loidl et al. \cite{GaHJH04}). Describing the temporal variability of various molecular line profiles visible in high-resolution spectra with the same models is another crucial test of whether  the models reproduce structures and dynamics of Mira atmospheres well. An important step is to reproduce simultaneously the observed behaviour of various types of lines (coming from regions with different velocities within the atmosphere) for several phases during the light cycle, and to compare synthetic RVs with observed ones.

For the line profile modelling discussed in this paper, a representative model atmosphere for a typical C-rich Mira (like e.g. S\,Cep) was calculated. The chosen  parameters are shown in Table\,\ref{t:dmaparameters}. The same model was also used for comparison with observed low-resolution IR spectra of S\,Cep in Gautschy-Loidl et al. (\cite{GaHJH04}) and can be found in their Table\,1 (l10t26c14u4f20pi). Figure\,\ref{f:structure} shows the corresponding spatial structures; the radius coordinate is plotted in units of the stellar radius of the initial model ($R_*$, calculated from $L_*$ and $T_*$). While the hydrostatic initial model is relatively compact, the atmosphere becomes much more extended as pulsation leads to a periodic levitation of the outer layers (compare Fig.\,2a of H\"ofner et al. \cite{HoGAJ03}). Shockfronts triggered by the pulsation propagate outwards. If temperatures drop below the dust condensation temperature and densities are high enough behind the shockfronts, dust can form (note that the dust shells in Fig.\,\ref{f:structure}c slightly lag behind the shocks in gas density seen in Fig.\,\ref{f:structure}b). Radiation pressure acting on the dust particles leads to an even more extended atmosphere. The dust shells move outwards, dragging along the gas by dynamic friction and a stellar wind develops. Figure\,\ref{f:structure} shows that the inner regions are dominated by the pulsation; the structures duplicate from one pulsational period to the next. The situation changes, as dust is being formed, starting at around 2\,R$_{*}$ (Fig.\,\ref{f:structure}c). As the time scales of pulsation and dust formation are  independent, the behaviour of these layers can differ greatly depending on the model parameters chosen. Dust shells are formed and evolve in different intervals (dust-induced $\kappa$-mechanism, cf. H\"ofner et al. \cite{HoeFD95} or Fleischer et al. \cite{FleGS95}) not necessarily periodically (compare Fig.\,1 of H\"ofner \& Dorfi \cite{HoefD97}) and the atmospheric structure is no longer repeating for the same phases of succesive periods. From $\approx$4--5\,R$_{*}$ outwards, the model shows a steady outflow with only small changes of the terminal velocity on time scales longer than the pulsational period. A transmitting outer boundary, fixed at 30\,R$_{*}$, is applied. Figure\,\ref{f:massenschalen} shows the temporal evolution of different layers.

Atmospheric structures from three different, separated cycles -- several snapshots from the temporal evolution of the dynamic model -- were used to calculate synthetic spectra. To denote model structures and synthetic spectra, bolometric phases $\phi_{\rm bol}$ within the lightcycle in luminosity will be used throughout the paper.

\begin{table}[h]
\begin{center}
\caption{Parameters and resulting properties of the dynamic model atmosphere used in this paper, chosen to resemble the C-rich Mira S\,Cep. The notation was adopted from H\"ofner et~al. (\cite{HoGAJ03}): 
$\Delta u_{\rm p}$ -- velocity amplitude of the piston at the inner boundary,
$\langle u \rangle$ -- mean outflow velocity at the outer boundary,
$\langle f_c$$\rangle$ -- mean degree of condensation of carbon into dust.}
\begin{tabular}{llc}
\hline
$L_*$&[L$_{\odot}$]&10$^4$\\
$M_*$&[M$_{\odot}$]&1.0\\
$T_*$&[K]&2600\\
$R_*$&[R$_{\odot}$]&493\\
&[AU]&2.29\\
lg $g_*$&&--0.94\\
C/O&&1.4\\
\hline
$P$&[d]&490\\
$\Delta u_{\rm p}$&[km\,s$^{-1}$]&4.0\\
$\Delta m_{\rm bol}$&[mag]&0.86\\
\hline
$\langle\dot M\rangle$&[M$_{\odot}\,$yr$^{-1}$]&4.3$\cdot$10$^{-6}$\\
$\langle u \rangle$&[km\,s$^{-1}$]&15\\
$\langle f_c$$\rangle$&&0.28\\
\hline
\end{tabular}
\label{t:dmaparameters}
\end{center}
\end{table}


\section{Spectral synthesis -- synthetic line profiles}     \label{s:synthesis}

In the following section, the basic properties of synthetic high-resolution spectra based on the dynamic model atmosphere described above are presented.


\subsection{Spectral lines, opacities and radiative transfer}
\label{s:linesopasST}

For the line modelling, several molecular and atomic lines were chosen according to the following criteria: (i) used successfully in published observations, (ii) line lists for the spectral synthesis are available, (iii) lines are observable from ground (atmospheric transmission), (iv) line depths, (v) low contamination by other lines or pseudo-continuous opacity. The high-resolution spectral atlases of Hinkle et al. (\cite{HinWL95}, \cite{HiWVH00}), Wallace \& Hinkle (\cite{WallH96}) and Ridgway et al. (\cite{RiCHJ84}) were useful for identifications. If necessary, in some cases these were also used to correct the positions of the molecular lines in the original line lists.

For the spectral synthesis, opacities calculated from line lists were used for CO (Goorvitch \& Chackerian \cite{GoorC94}) and CN (J{\o}rgensen \& Larsson \cite{JorgL90}), as described in Aringer et al. (\cite{AriKL97}, \cite{AriKJ02}). The spectra were computed with a resolution of $\lambda/\Delta\lambda$=300\,000 and then rebinned to 70\,000, which is comparable to observed FTS spectra. For the calculations we adopted Doppler profiles assuming a microturbulence velocity of 2.5\,km\,s$^{-1}$, which is consistent with previous modelling (e.g. Aringer \cite{AHWHJ99}) and should have no major impact on the computations as it is small compared to the macroscopic gas velocities of the model.

For consistency and in order to get realistic spectra, molecules other than CO and CN should be accounted for in the radiative transfer as well. Also the optical depth at which the lines originate should be as realistic as possible. Thus, the line list data for CO and CN was supplemented with opacity sampling (OS) data for other C-bearing molecules (as described in detail in Gautschy-Loidl et al. \cite{GaHJH04}), given with much lower spectral resolution than that  used in the computation of the line profiles. In order to account for the quasi-continuous opacity contribution of these species, one opacity value\footnote{Constant over wavelength for all layers, but variable as a function of depth.} -- averaged from sufficient ($\approx$300) OS data points around the central wavelength of the considered line -- was added to each wavelength point of the high-resolution spectra calculated with line lists. It should be emphasised that the method works only for molecules with many densely-spaced absorption lines affecting the spectra more or less continuously ("pseudo-continuum"). As the absorption changes only slightly and a mean value is larger than the fluctuations, averaging delivers meaningful numbers.\footnote{For other molecules (like C$_2$), the described method results in a large overestimation of the opacity, since opacity distributions with pronounced features must not be averaged. Opacities therefore cannot be approximated by the discussed pseudo-continuum approach.} We found that the only relevant molecule was C$_2$H$_2$.

Furthermore, frequency-dependent opacities of C-rich dust (pure amorphous carbon, no SiC) were included using the data (set AC) of Rouleau \& Martin (\cite{RoulM91}). For the moderately varying dust opacity, one mean, constant value was added at each wavelength point of the high-resolution spectra computed from line lists.

The radiative transfer code which was adapted and used for the spectral synthesis is described in detail in Windsteig (\cite{Winds98}); we only give a short overview here. The code solves radiative transfer in spherical geometry by integrating the frequency-dependent transfer equation along characteristic rays (a description of the applied method can be found in Yorke \cite{Yorke88}). In the current version, velocity effects are also accounted for. Because of the diverse movements of atmospheric layers at different depths, including the influence of velocities on the interaction between matter and radiation is by definition essential to model the complex behaviour of spectral lines in Mira spectra. For all results presented here LTE conditions were assumed.\footnote{This assumption may not be justified for layers crossed by a shock wave and in the extended regions some $R_*$ away from the star. This has been explored e.g. by Schirrmacher et al. (\cite{SchWS03}), who state that non-LTE effects "are generally more pronounced outside of the dust formation zone, where the densities are smaller". It is supposed that these effects do not contribute to the studies presented here (also assumed by Winters et al. \cite{WiKGS00}).} Level populations were computed from Boltzmann distributions at the corresponding gas temperature (resulting from the non-grey radiative transfer of the atmospheric modelling) and then used to solve the observer's frame radiation transport equation by means of a Rybicki scheme. The spatial resolution of the radiative transfer (radial points and impact parameters) is determined by the mesh points of the modelling and therefore temporally varying due to the adaptive grid used. Additional depth points are inserted in the case of large radial (velocity) gradients. First results obtained with this code based on dynamic models were presented by Windsteig et al. (\cite{WiHAD98}, \cite{WiLHD98}, \cite{WiALH99}) and Nowotny et al. (\cite{NAGHL03}, \cite{NAHGH04}).

All synthetic spectra are normalised relative to a computation where only continuum opacity is taken into account (no contribution of the molecular line itself).


\subsection{Naming convention for velocities}
\label{s:velocityconvention}

In all observational papers, velocities are defined from the observer's point of view, which by convention is positive for material moving away from the observer  and negative for matter moving towards the observer. Thus blue-shifted absorption lines -- indicating outflow from the star -- give negative RVs, while infalling matter revealed by red-shifted lines results in positive RVs. For easier comparison, this convention was adopted for the plots here (e.g. Fig.\,\ref{f:structure}). Line profiles are then plotted with RV as abscissa, computed from the central wavelength and by using the formula for Doppler shift.


\subsection{Influence of velocities on optical depth}
\label{s:influence}

\begin{figure}
   \resizebox{\hsize}{!}{\includegraphics{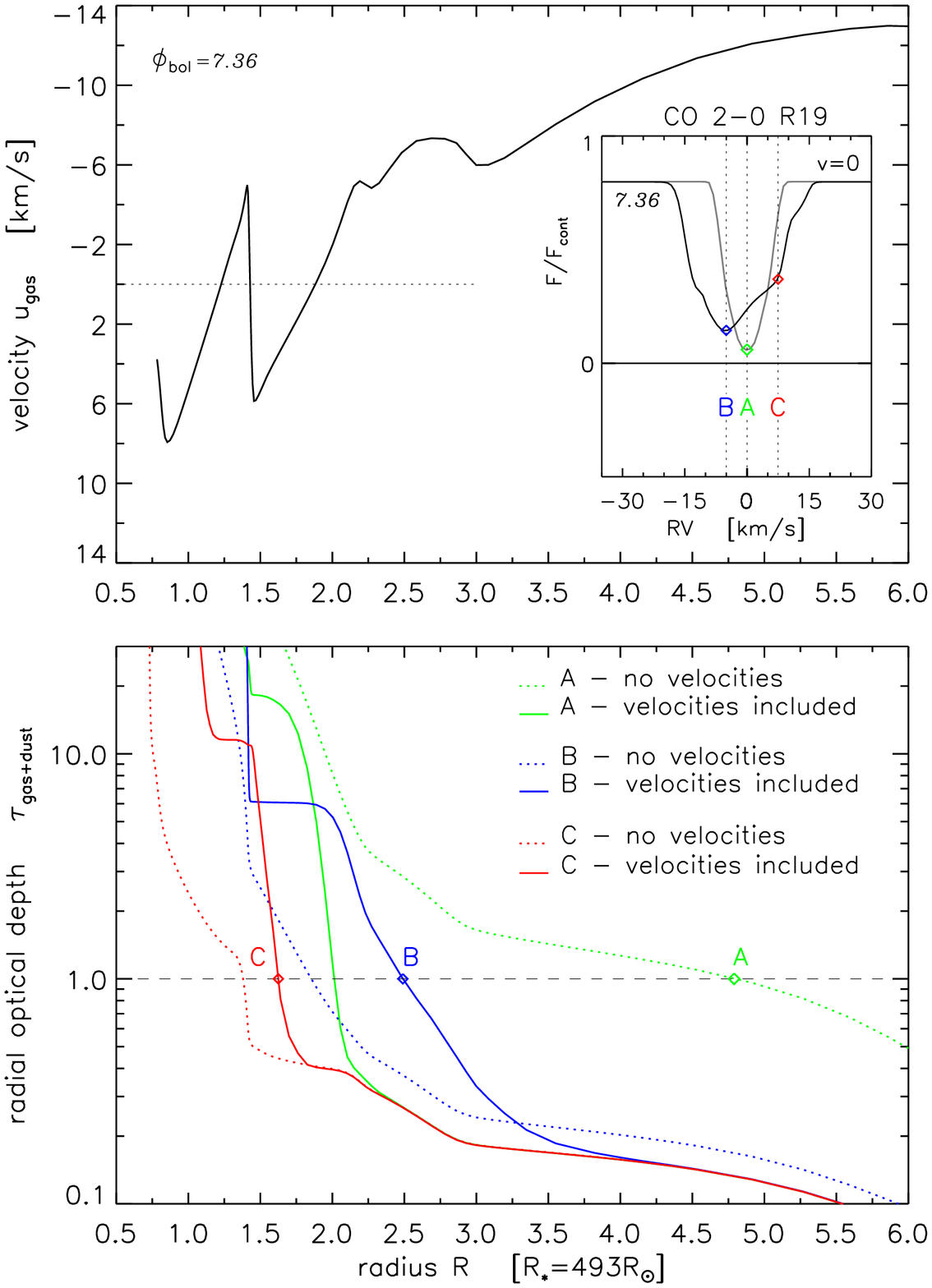}}
    \caption[]
    {\textit{Upper panel:} Gas velocities at a certain phase and synthetic
    CO $\Delta v$=2
    line profiles calculated with/without taking velocities into account in
    the radiative transfer.\\
    \textit{Lower panel:} Demonstration of how velocities in the atmosphere
    change optical depths and the approximate region at which the line
    originates (indicated by the $\tau$=1--line).}
   \label{f:co2muetauplot}
\end{figure}

Figure\,\ref{f:co2muetauplot} demonstrates for the model used how velocities within Mira atmospheres can influence the process of line formation, by considering the CO 2--0 R19 low-excitation line (cf. Sect.\,\ref{s:co1stovertone}) as an example. In the lower panel dotted lines show how the values of radial optical depth increase inwards by calculating $\tau$ from radial integration over opacities and densities for the rest wavelength in the line center (A) and two wavelength points in the line wings (B, C -- see insert in the upper panel of the figure). Solid lines show the corresponding optical depths if the opacities in all layers of the model atmosphere are Doppler-shifted according to the gas velocity there. 

For the rest wavelength of the line (corresponding to $RV$=\,0\,km\,s$^{-1}$ -- A), a large fraction of the opacity is shifted away due to the outflow in the outer regions. Therefore $\tau$=1 is reached much deeper in the atmosphere if velocities are considered. It is the other way round for the two wavelength points in the line wings marked B and C. The locations where optical depths reach unity move further away from the star, because additional opacity is moved to these wavelengths by Doppler shift. The optical depth for B is almost constant between 1.5--2\,R$_*$, as most of the opacity is red-shifted by strong infall. This redistribution of opacity results in a broadened line, with a blue-shifted line minimum. The strong influence of velocities on the region of line formation is quite evident. If $\tau$=1 is considered to measure the approximate location of the line-forming region, this is shifted from $\approx$5\,R$_*$ [A] to  $\approx$2.5\,R$_*$ [B] (corresponding to a difference of $\approx$5.73\,AU).

The low-excitation CO $\Delta v$=2 lines originate from atmospheric layers with a wide range of velocities, hence a broad profile, and large changes in $\tau$ result. Other lines, e.g. CO $\Delta v$=3, originate in a relatively narrow depth range with more uniform velocities. Therefore, the shifted and split components show roughly the same line width as profiles computed without velocities (Fig.\,\ref{f:co16mueprofiles}), and the difference of the position of $\tau$=1 is smaller if velocities are taken into account.


\subsection{Probing the pulsational layers}  \label{s:pulslayers}


\subsubsection{CO $\Delta v$=3 lines}  \label{s:co2ndovertone}

Around 1.6$\mu$m the continuous opacity has a minimum (H$^{-}$--peak), allowing us to see deep into the atmosphere. This, in combination with the large excitation energies of second overtone vibration-rotation lines of CO at these depths, led to the fact that these lines have been most used to sample regions where motions are dominated by pulsation of the stellar interior (HSH84, Lebzelter \& Hinkle \cite{LebzH02}). Figure\,1 of HHR82 shows a clear demonstration of the very characteristic behaviour for the S-type Mira $\chi$\,Cyg by means of averaged line profiles for different phases. Coupled to the lightcycle, periodic variations in wavelength shift can be seen, which repeat in the same way every pulsational period. A blue-shifted component appears around $\phi_{\rm v}$$\approx$0.9, becomes stronger, moves towards red-shifts, becomes weaker and disappears approximately one period later at $\phi_{\rm v}$$\approx$0.1. For a certain time interval around phases of visual maximum, the lines appear doubled. This is due to the fact that infalling material (red-shifted) can be observed together with gas being accelerated outwards by the next emerging shock wave (blue-shifted component) at the same time (compare Fig.\,1 of Alvarez et al. \cite{AJPGF00}). This behaviour results in \mbox{S-shaped,} discontinuous RV-curves as in Fig.\,12 of HHR82 and is usually explained by radial pulsations and emerging shockfronts.

\begin{figure}
   \resizebox{\hsize}{!}{\includegraphics{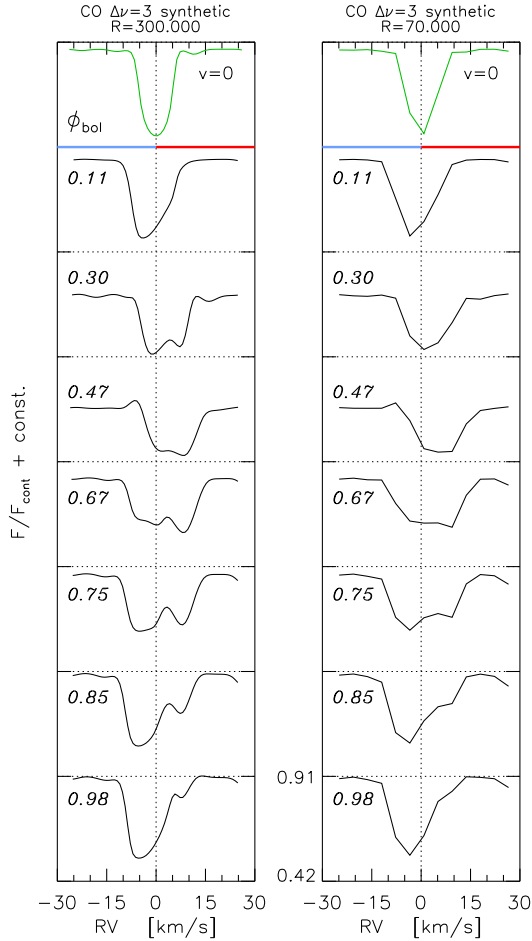}}
    \caption[]
    {Time series of synthetic CO 5--2 P30 line profiles during one lightcycle
    with different spectral resolutions.}
   \label{f:co16mueprofiles}
\end{figure}  

Since observing second overtone CO lines is difficult for C stars due to contamination of this spectral region by other molecules (mainly CN, C$_2$), no corresponding studies have yet been published. Nevertheless, we synthesized such line profiles, since observational studies indicate a very similar dynamical behaviour for spectral types M, S and C (compare Lebzelter \& Hinkle \cite{LebzH02}). 

The CO 5--2 P30 line at 6033.8967\,cm$^{-1}$ (1.6573$\mu$m) was chosen for modelling. CN and C$_2$, which are also prominent in this spectral region, were not taken into account. Including opacities from line lists of these lines could (due to uncertainties in positions) influence the profiles by blending.\footnote{Only from synthetic spectra of this spectral region, the CO \mbox{5--2} P30 line appears to be in a "window" of the CN-line-forest. This may allow a comparison with observed high-resolution spectra in the future.} This specific synthetic CO line may then not be directly comparable to the corresponding observed (and by CN/C$_2$ influenced) one. But since all CO $\Delta v$=3 lines in this region have similar excitation potentials and gf values, the synthetic line should be comparable to an observed average line profile (where the various influences smooth out). The synthetic spectra here are computed in a slightly different way than the preliminary results presented and analysed in Nowotny et al. (\cite{NAHGH04}). Dust opacities are from Rouleau \& Martin (\cite{RoulM91}) instead of Maron (\cite{Maron90}). The former contains more complete data over a larger spectral range and was also used for the modelling of the atmospheric structures. In addition, the pseudo-continuous opacity of C$_2$H$_2$ is also included here, leading to smaller line depth due to a depressed "continuum".

Figure\,\ref{f:co16mueprofiles} shows time series of the synthetic second overtone CO line profiles for two different spectral resolutions. To get an idea of the typical line width, a spectrum computed for the phase of light maximum and without taking velocities into account in the radiative transfer is plotted at the top. Due to the limited number of spectral points the profil looks asymmetric and not centered on the rest wavelength ($RV$=0) at lower resolution. In principle, the above-described typical observed behaviour of CO $\Delta v$=3 lines in M/S stars can be recognised from the synthetic spectra.

Some deviations from observations of Mira variables can be found. Most noticeable is the fact that the transition from blue- to red-shift during phases $\phi_{\rm bol}$$\approx$\textit{0.2--0.5} is only visible in the lower-resolution spectra. The line profiles look more complicated at higher spectral resolutions. There, the movement of the original component stops at $RV$$\approx$\,0\,km\,s$^{-1}$ where it disappears, while another red-shifted component develops from $\phi_{\rm bol}$$\approx$\textit{0.2} on. On the other hand, line splitting is more pronounced at higher resolution ($\phi_{\rm bol}$=\textit{0.75}). From the synthetic spectra it can be noticed that the (pseudo-) continuous opacity is strongest\footnote{Around minimum phase, the rate of dust production is highest and also the C$_2$H$_2$ features are strongest for a dusty model like the one used here.} for $\phi_{\rm bol}$$\approx$\textit{0.4}, leading to a depressed "continuum" and apparently weaker lines. A more detailed analysis shows that the synthetic CO $\Delta v$=3 lines emerge from depths of $R$=0.8--1.3\,R$_*$ with gas temperatures of $\approx$2200--3500\,K. This seems consistent with the observational results of HHR82 (their Fig.\,5) or HSH84, listing excitation temperatures of $\approx$2000--4500\,K. As it would be expected for lines coming from the very inner parts of the atmosphere with periodic movements (Fig.\,\ref{f:massenschalen}), line shapes are reproduced for all three chosen periods. Therefore only profiles from the first one are shown.

\subsubsection{CN $\Delta v$=--2 lines}   \label{s:cnprofiles}

From the visual range to beyond 2$\mu$m, spectra of C stars are dominated by features of CN and C$_2$. Coming from deep photospheric layers (Fig.\,3 of Loidl et al. \cite{LoHJA99}), CN lines are also suited for investigating kinematics in pulsational layers, as it has been shown by HB96. HB96 used lines around 2.14$\mu$m, which are electronic transitions of the red system of CN with $\Delta v$=--2. These lines show the same typical behaviour as found for the CO $\Delta v$=3 lines (Sect.\,\ref{s:co2ndovertone}), also leading to a discontinuous RV-curve.

\begin{figure}
   \resizebox{\hsize}{!}{\includegraphics{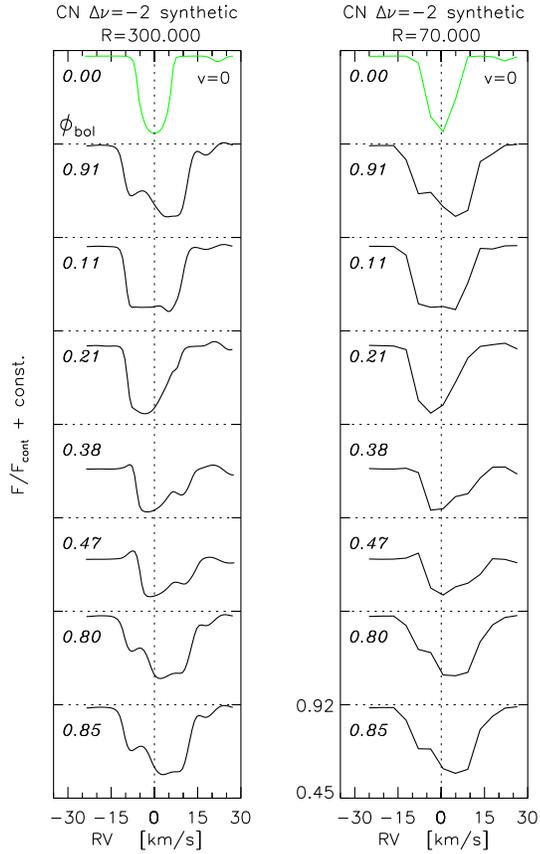}}
    \caption[]
    {Evolution of synthetic line profiles for the CN 
    1--3 Q$_2$4.5 line during a pulsational period at higher and 
    lower resolution.}
   \label{f:cnprofiles}
\end{figure}

For the spectral synthesis the 1--3 Q$_2$4.5 line at 4871.340\,cm$^{-1}$ (2.0528$\mu$m) was chosen. The influence of C$_2$H$_2$ is rather weak there, but observations are difficult due to telluric contamination. No molecules other than C$_2$H$_2$ are relevant for the overall continuous opacity in this region. As in the case of CO, the behaviour of this CN line should be similar to those typically used in observational studies.

Figure\,\ref{f:cnprofiles} shows the results of the spectral synthesis for several phases during the lightcycle. As they are the same from one period to the next, spectra are again shown only for one lightcycle. Observed profile variations can qualitatively be reproduced by the calculations. The pattern with blue-/red-shifts and line doubling around light maximum (Sect.\,\ref{s:co2ndovertone}) is discernible. However, the line shapes appear somewhat complex. The splitting is less pronounced than for the CO $\Delta v$=3 lines; the two components are visible but merge. The strength of the red component does not decrease continuously. Several components are visible at higher resolution (e.g. $\phi_{\rm bol}$=\textit{0.80}). This is better seen by rebinning the synthetic spectra down to resolutions comparable to observed FTS spectra (70\,000). The line profiles are smoothed and especially the transition from blue- to red-shift during phases of $\approx$\textit{0.2--0.8} appears more distinct. While for example at phase \textit{0.80} the two visible red-shifted components  merge into one broad feature, it becomes difficult to measure the weak blue-shifted component. The \mbox{(pseudo-)} continuous opacity (C$_2$H$_2$, dust) is again strongest for $\phi_{\rm bol}$$\approx$\textit{0.4}.

With line profiles being even more complex, CN lines qualitatively show the same behaviour as CO $\Delta v$=3 lines. But -- still sampling deep layers driven by pulsation -- they originate from slightly smaller optical depths. This can be deduced from line doubling at later phases.

In addition, a CN $\Delta v$=--1 line of the 0--1 red system in the H-band was calculated. This 0--1 Q$_1$76.5 line at 6107.012\,cm$^{-1}$ (1.6374$\mu$m) shows almost the same behaviour (profiles, RVs) as described above, being only different in a very small shift in $\phi_{\rm bol}$. This was also found for a Ti line in the K-band where observed line doubling around light maximum (HB96) could clearly be reproduced.


\subsection{Probing the dust-forming region -- CO $\Delta v$=2:}
\label{s:co1stovertone}

The first overtone vibration-rotation lines of CO in the K-band around 2$\mu$m are also clearly observable. These have been studied in time series of spectra for several stars: for the C-rich Mira S\,Cep by HB96, for the C-rich Mira IRC+10216 by Winters et al. (\cite{WiKGS00}), for the S-type Mira $\chi$\,Cyg by HHR82, for the O-rich Mira R\,Leo by Hinkle (\cite{Hinkl78}) and other M stars e.g. by Lebzelter et al. (\cite{LebHH99}).

\begin{figure}[h]
   \resizebox{\hsize}{!}{\includegraphics{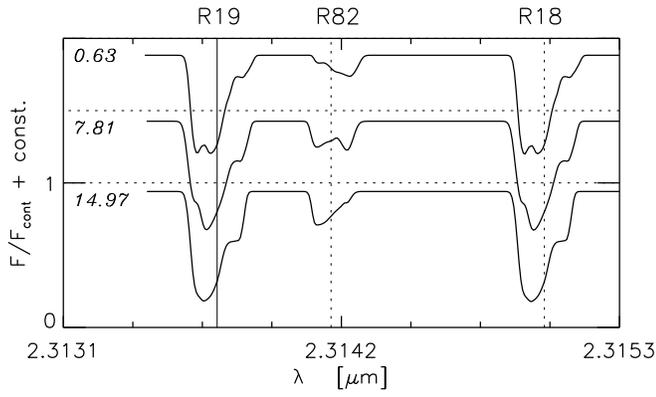}}
    \caption[]
    {Synthetic spectra (resolution of $R$=300\,000) containing CO $\Delta v$=2
    lines and demonstrating the different behaviour of 
    low (e.g. 2--0 R19) and high excitation lines 
    (e.g. 2--0 R82). See text for details. 
    Marked are the rest wavelengths of the lines.}
   \label{f:co2muelowhigh}
\end{figure}

For the line modelling a few suitable lines used in these studies were chosen, especially the CO 2--0 R19 low-excitation line at 4322.0657\,cm$^{-1}$ (2.3137$\mu$m) and the CO 2--0 R82 high-excitation line at 4321.2240\,cm$^{-1}$  (2.3142$\mu$m). Concerning the treatment of other molecules, the statements of Sect.\,\ref{s:co2ndovertone} are also valid here. Figure\,\ref{f:co2muelowhigh} shows a few synthetic spectra based on our dynamic model.

The behaviour of CO $\Delta v$=2 lines is two-fold. The weaker high-excitation lines (e.g. R82 in Fig.\,\ref{f:co2muelowhigh}) act in the same way as second overtone lines and almost duplicate their S-shaped RV-curve. While this is clearly visible for M- (Hinkle \cite{Hinkl78}) or S-type (HHR82) stars, contamination by many other lines (mainly CN) in this spectral region hampers the line identification for C-rich stars (see HB96). Figure \ref{f:co2muelowhigh} shows that our model can reproduce the scenario for high-excitation lines reasonably well (compare Fig.1 of Alvarez et al. \cite{AJPGF00}).

\begin{figure}
   \resizebox{\hsize}{!}{\includegraphics{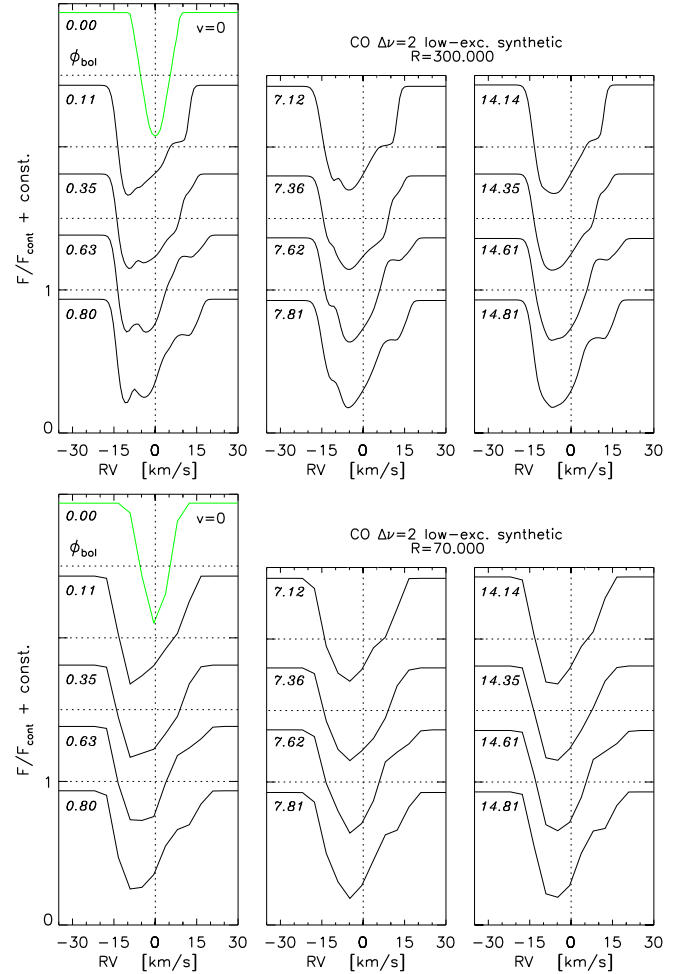}}
    \caption[]
    {Comparison of synthetic line profiles of the CO first overtone 
    low-excitation line 2--0 R19 for comparable phases from different 
    lightcycles. Substructures that are visible at higher spectral resolution 
    (\textit{upper panel}) are smoothed out at lower resolution 
    (\textit{lower panel}).}
   \label{f:co2mueprofiles}
\end{figure}

In the following, only the stronger low-excitation lines (e.g. R19 in Fig.\,\ref{f:co2muelowhigh}) will be considered. Easier to identify in the spectra, they show a somewhat different behaviour. Variability is less pronounced than for second overtone CO lines (which appear shifted/split with approximately the same line width). They show complex, asymmetric shapes and seem to consist of several components, which are not further separable (Fig.\,3 in HB96). This is reproduced by our model; Fig.\,\ref{f:co2mueprofiles} presents a series of line profiles. A comparison with the profile calculated without taking velocities into account suggests that various layers with different velocities contribute to the final, broadened shapes. The variation is not repeating for the same phase of different lightcycles (see Sect.\,\ref{s:modelling}). Again the pseudo-continuous opacity (C$_2$H$_2$, dust) is strongest for $\phi_{\rm bol}$$\approx$\textit{0.4}.

Observed radial velocities being more or less constant around the CMRV leads to the scenario of a static layer within Mira atmospheres (e.g. HHR82 or Tsuji\,\cite{Tsuji88}) -- cf. Nowotny et al. (\cite{NowLH05}, Paper\,II) for a more detailed discussion. Also for the synthetic line profiles in Fig.\,\ref{f:co2mueprofiles}, most of the time one blue-shifted main component around $\approx$5\,km\,s$^{-1}$ is visible, which would be expected from the estimated region of line formation (Fig.\,\ref{f:co2muetauplot} -- B) and the velocities there (Fig.\,\ref{f:regions}). In these depths of $\approx$2--3\,R$_*$, gas temperatures of $\approx$800--1500\,K are found. This seems consistent with the excitation temperatures of 800$\pm$100\,K derived for the same lines in  $\chi$\,Cyg spectra by HHR82. Low-excitation first overtone CO lines can be used to probe dynamics in layers where dust is being formed and the outflow starts (Fig.\,\ref{f:structure}cd). Similar conclusions were also drawn by Winters~et~al. (\cite{WiKGS00}).


\subsection{Probing the outflow -- CO $\Delta v$=1:}

Finally we study the fundamental mode vibration-rotation lines of CO located at $\approx$4$\mu$m. Not much observational material exists, as observations in this spectral region are hampered by telluric absorption. Nevertheless, these lines are useful as they sample the outermost layers of Mira atmospheres. Results on this have been presented for $\chi$\,Cyg by HHR82, for IRC +10216 (CW\,Leo) by Keady et al. (\cite{KeaHR88}) and for $\alpha$ Ori by Bernat et al. (\cite{BeHHR79}).

\begin{figure}[h]
   \resizebox{\hsize}{!}{\includegraphics{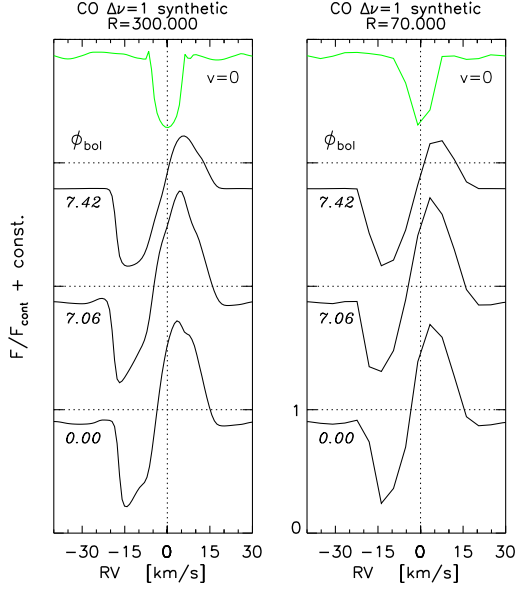}}
    \caption[]
    {Synthetic CO fundamental lines 1--0 R1, showing characteristic 
    \mbox{P Cygni-type} profiles
     indicative of a stellar wind.}
   \label{f:co4mueprofiles}
\end{figure}

The CO 1--0 R1 line at 2150.856\,cm$^{-1}$ (4.6493$\mu$m) was chosen for modelling. Only the pseudo-continuous opacity of C$_2$H$_2$ and dust is taken into account, since no other molecules contribute significantly in this region. Figure\,\ref{f:co4mueprofiles} shows a time series of synthetic spectra compared to one calculated without velocities taken into account. The line profile looks qualitatively the same for all phases of the model and shows a typical P\,Cygni-type shape with a deep, blue-shifted absorption component (from the outflowing material in the line of sight) and a superimposed red-shifted emission component (from the extended regions around the star). The strength of the emission is variable; in Fig.\,\ref{f:co4mueprofiles} the extreme cases are shown. Similar synthetic profiles were presented by Winters et al. (\cite{WiKGS00}) and compared to the observed ones of Keady et al. (\cite{KeaHR88}). From radial gradients of $\tau$ (as Fig.\,\ref{f:co2muetauplot}) it is found that these lines originate as expected in the wind region at $\approx$15\,R$_*$. Low gas temperature of $\approx$350--500\,K in the line forming regions there appear comparable with excitation temperatures of 300$\pm$200\,K given by HHR82.

\begin{figure}[h]
    \resizebox{\hsize}{!}{\includegraphics{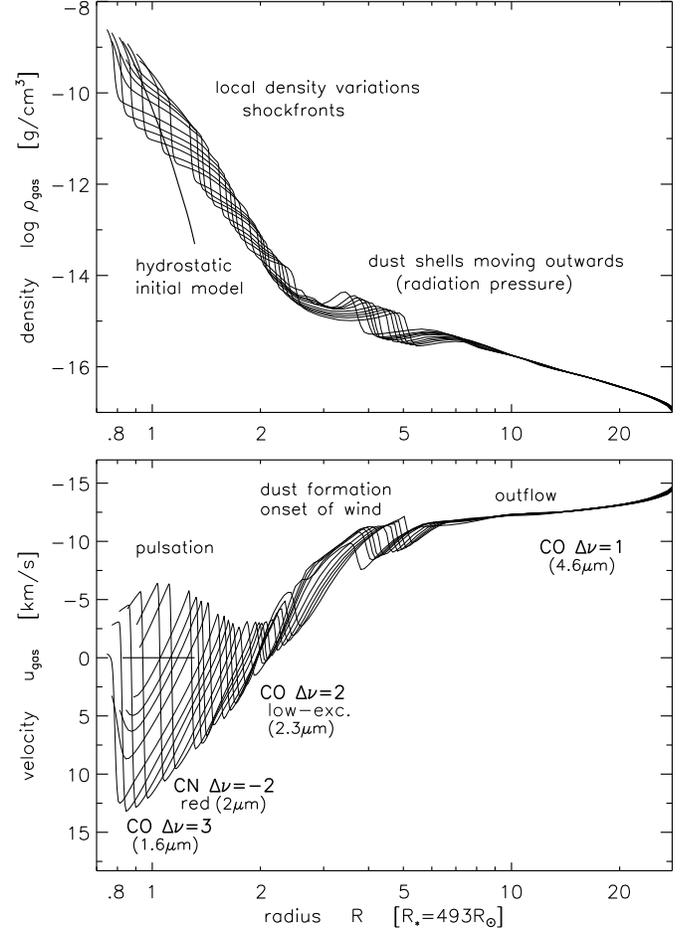}}
    \caption[]
    {\textit{Upper panel:} Density structures of the dynamic model 
    atmosphere for  
    several phases during 
    one pulsational cycle ($\phi_{\rm bol}$=\textit{0.0--1.0}), 
    demonstrating the larger extension
    compared to the hydrostatic initial model and the density variations --
    shockfronts due to pulsation in the inner regions and dust shells
    propagating outwards further out.\\
    \textit{Lower panel:} Corresponding velocity structures; 
    approximate regions probed by different types of molecular lines
    are marked,
    inferred from the velocity behaviour and plots of the optical depths.}
   \label{f:regions}
\end{figure}


\section{Summary and outlook}

Extended AGB star atmospheres are complicated due to different interacting physical processes such as pulsation, formation of molecules and dust and eventually the developement of mass loss. This leads to complicated, non-monotonic velocitiy fields. From several studies in the past it is known that different molecular lines in high-resolution IR spectra are useful for probing kinematics throughout these Mira atmospheres. To simulate and understand the processes taking place there, dynamic model atmospheres are constructed. They are especially needed to reproduce variable phenomena, one of the most crucial aspects being the temporal variation of line profiles. 

Using the model atmospheres of H\"ofner et al. (\cite{HoGAJ03}) we were able to reproduce the typical global velocity structure (derived from Mira observations) from the pulsating atmospheric layers through the dust-forming region out to the outflow region (Fig.\,\ref{f:regions}). It was demonstrated that lines originating in layers of different depths can be modelled simultaneously by only one consistently computed atmospheric model and successive radiative transfer. It is absolutely necessary to include velocity effects in the radiative transfer to model the complex line profiles and their variations.

The behaviour of lines sampling different regions, as known from observations, could qualitatively be reproduced, as summarised in Fig.\,\ref{f:regions}:
\begin{itemize}
\item
CO $\Delta v$=3 lines, CO $\Delta v$=2 high-excitation lines and CN $\Delta v$=--2 lines sample the deep photosphere, which is dominated by pulsation. At low spectral resolution (70\,000), time series of synthetic line profiles resemble observations (compare Fig.\,1 of HHR82), although some differences remain (e.g. line doubling is significantly too weak). At highest resolution (300\,000), the profiles appear more complex. It will be interesting to see if future instruments (with higher resolutions than the FTS spectrograph used by Hinkle and collaborators) will allow us to verify whether there are such substructures in the profiles or if this is a pecularity of the model atmospheres used.

\item
CO $\Delta v$=2 low-excitation lines sample the dust-forming region, where the stellar wind is triggered. They appear therefore slightly blue-shifted (reminiscent of a static layer) and show complex, broadened shapes with irregular temporal variation.

\item
CO $\Delta v$=1 lines probe the layers of steady outflow and show typical P\,Cygni-type shapes at any time.

\end{itemize}

Thus, it can be ascertained that the dynamic model atmosphere used here shows fundamental agreement with dynamic processes (pulsation, dust formation, mass loss mechanism) occuring in Mira atmospheres.

A detailed comparison of the synthetic spectra with observed FTS spectra of the carbon-rich Mira S\,Cep and radial velocities derived from line profiles is presented in Nowotny et al. (\cite{NowLH05}, Paper\,II).


\begin{acknowledgements}
WN wishes to thank Kjell Eriksson and the Uppsala Observatory staff for their support (scientifically and financially) during his stays at Uppsala. Sincere thanks are given to T. Lebzelter and J. Hron for careful reading and fruitful discussions. This work was supported by the ``Fonds zur F\"orderung der Wis\-sen\-schaft\-li\-chen For\-schung'' under project number P14365--PHY and the Swedish Research Council. 

\end{acknowledgements}


\begin{thebibliography}{}

\bibitem [2000] {AJPGF00}
Alvarez, R., Jorissen, A., Plez, B., et al. 2000, A\&A, 362, 655
\bibitem [2001a] {AJPGF01a}
Alvarez, R., Jorissen, A., Plez, B., et al. 2001a, A\&A, 379, 288
\bibitem [2001b] {AJPGF01b}
Alvarez, R., Jorissen, A., Plez, B., et al. 2001b, A\&A, 379, 305
\bibitem [1997] {AriKL97}
Aringer, B., J{\o}rgensen, U.G., \& Langhoff, S.R. 1997, A\&A, 323, 202
\bibitem [1999] {AHWHJ99}
Aringer, B., H\"ofner, S., Wiedemann, G., et al. 1999, A\&A, 342, 799
\bibitem [2002] {AriKJ02}
Aringer, B., Kerschbaum, F., \& J{\o}rgensen, U.G. 2002, A\&A, 395, 915
\bibitem [1992a] {Barnb92a}
Barnbaum, C. 1992a, ApJ, 385, 694
\bibitem [1992b] {Barnb92b}
Barnbaum, C. 1992b, AJ, 104, 1585
\bibitem [1995] {BarnH95}
Barnbaum, C., \& Hinkle, K.H. 1995, AJ, 110, 805
\bibitem [1979] {BeHHR79}
Bernat, A.P., Hall, D.N.B., Hinkle, K.H., \& Ridgway, S.T. 1979, ApJ, 233, L135
\bibitem [1988] {BesBS88}
Bessell, M.S., Brett, J.M., \& Scholz, M., et al. 1988,
in Atmospheric Diagnostics of Stellar Evolution: Chemical Pecularity, Mass Loss, and Explosion, IAU Coll. 108, ed. K. Nomoto, Springer, p.187
\bibitem [1989] {BeBSW89a}
Bessell, M.S., Brett, J.M., Scholz, M., \& Wood, P.R. 1989, A\&A, 213, 209
\bibitem [1989] {BessS89}
Bessell, M.S., \& Scholz, M. 1989, 
in Evolution of Peculiar Red Giants, IAU Coll. 106, ed. H.R. Johnson, B. Zuckerman, Cambridge University Press, p.67
\bibitem [1996] {BesSW96}
Bessell, M.S., Scholz, M., \& Wood, P.R. 1996, A\&A, 307, 481
\bibitem [1999] {BusGW99}
Busso, M., Gallino, R., \& Wasserburg, G.J. 1999, ARA\&A 37, 239
\bibitem [1995] {FleGS95}
Fleischer, A.J., Gauger, A., \& Sedlmayr, E. 1995, A\&A, 297, 543
\bibitem [1990] {GauGS90}
Gauger, A., Gail, H.-P., \& Sedlmayr, E. 1990, A\&A, 235, 345
\bibitem [2004] {GaHJH04}
Gautschy-Loidl, R., H\"ofner, S., J{\o}rgensen, U.G., \& Hron, J. 2004, A\&A, 422, 289
\bibitem [1994] {GoorC94}
Goorvitch, D., \& Chackerian, C.Jr. 1994, ApJS, 91, 483
\bibitem [2004] {GustH04}
Gustafsson, B., \& H\"ofner, S.
2004, in Asymptotic Giant Branch Stars, ed. H.J. Habing, H. Olofsson, Springer, ch.4, p.149
\bibitem [1979] {HillW79}
Hill, S.J., \& Willson, L.A. 1979, ApJ, 229, 1029
\bibitem [1978] {Hinkl78}
Hinkle, K.H. 1978, ApJ, 220, 210
\bibitem [1979a] {HinkB79a}
Hinkle, K.H., \& Barnes, T.G. 1979a, ApJ, 227, 923
\bibitem [1979b] {HinkB79b}
Hinkle, K.H., \& Barnes, T.G. 1979b, ApJ, 234, 548
\bibitem [1996] {HinkB96}
Hinkle, K.H., \& Barnbaum, C. 1996, AJ, 111, 913 (HB96)
\bibitem [1982] {HinHR82}
Hinkle, K.H., Hall, D.N.B., \& Ridgway, S.T. 1982, ApJ, 252, 697 (HHR82)
\bibitem [1984] {HinSH84}
Hinkle, K.H., Scharlach, W.W.G., \& Hall, D.N.B. 1984, ApJ Suppl., 56, 1 (HSH84)
\bibitem [1995] {HinWL95}
Hinkle, K., Wallace, L., \& Livingston, W. 1995,
``Infrared Atlas of the Arcturus Spectrum, 0.9--5.3 microns'' (San Francisco: Astronomical Society of the Pacific)
\bibitem [2000] {HiWVH00}
Hinkle, K., Wallace, L., Valenti, J., \& Harmer, D. 2000,
``Visible and Near Infrared Atlas of the Arcturus Spectrum, 3727--9300\,\AA'' (San Francisco: Astronomical Society of the Pacific)
\bibitem [1983] {Hirai83}
Hirai, M. 1983, in The MK Process and Stellar Classification, ed. R. Garrison, David Dunlap Observatory, p.353
\bibitem [1998] {HofSW98}
Hofmann, K.-H., Scholz, M., \& Wood, P.R. 1998, A\&A, 339, 846
\bibitem [1999] {Hoefn99}
H\"ofner, S. 1999, in Asymptotic Giant Branch Stars, ed. T. Le Bertre, A. Lebre, C. Waelkens, IAU Symp. 191, p.159
\bibitem [1997] {HoefD97}
H\"ofner, S., \& Dorfi, E.A. 1997, A\&A, 319, 648
\bibitem [1995] {HoeFD95}
H\"ofner, S., Feuchtinger, M.U., \& Dorfi, E.A. 1995, A\&A, 297, 815
\bibitem [2003] {HoGAJ03}
H\"ofner, S., Gautschy-Loidl, R., Aringer, B., \& J{\o}rgensen, U.G. 2003, 
A\&A, 399, 589
\bibitem [2004] {HGANH04}
H\"ofner, S., Gautschy-Loidl, R., \& Aringer, B., et al. 2004,
in Proc. of ESO Workshop "High-resolution IR spectroscopy in Astronomy", ed. H.U. K\"aufl, R. Siebenmorgen, A. Moorwood, ESO Astrophysics Symposia, Springer, p.279
\bibitem [2003] {JeWLS03}
Jeong, K.S., Winters, J.M., Le Bertre, T., \& Sedlmayr, E. 2003, A\&A, 407, 191
\bibitem [2000] {JorHL00}
J{\o}rgensen, U.G., Hron, J., \& Loidl, R. 2000, A\&A, 356, 253
\bibitem [1990] {JorgL90}
J{\o}rgensen, U.G., \& Larsson, M. 1990, A\&A, 238, 424
\bibitem [1954] {Joy54}
Joy, A.H. 1954, ApJ Suppl., 1, 39
\bibitem [1988] {KeaHR88}
Keady, J.J., Hall, D.N.B., \& Ridgway, S.T. 1988, ApJ, 326, 832
\bibitem [2002] {LebzH02}
Lebzelter, T., \& Hinkle, K.H. 2002, 
in Radial and Nonradial Pulsations as Probes of Stellar Physics, IAU Coll. 185,
ed. C.\,Aerts, T.R.\,Bedding, J.\,Christensen-Dalsgaard, ASP Conf. Series, 259, 556
\bibitem [1999] {LebHH99}
Lebzelter, T., Hinkle, K.H., \& Hron, J. 1999, A\&A, 341, 224
\bibitem [2001] {LebHA01}
Lebzelter, T., Hinkle, K.H., \& Aringer, B. 2001, A\&A, 377, 617
\bibitem [1999] {LoHJA99}
Loidl, R., H\"ofner, S., J{\o}rgensen, U.G., \& Aringer, B. 1999, A\&A, 342, 531
\bibitem [2001] {LoiLJ01}
Loidl, R., Lancon, A., \& J{\o}rgensen, U.G. 2001, A\&A, 371, 1065
\bibitem [1990] {Maron90}
Maron, N. 1990, Ap\&SS, 172, 21
\bibitem [2004] {Milla04}
Millar, T.J. 2004, in Asymptotic Giant Branch Stars, ed. H.J. Habing, H. Olofsson, Springer, ch.5, p.247
\bibitem [2003] {MolsW03}
Molster, F.J., \& Waters, L.B.F.M. 2003, 
in Astromineralogy, ed. Th. Henning, Springer, p.121
\bibitem [2003] {NAGHL03}
Nowotny, W., Aringer, B., Gautschy-Loidl, R., et al. 2003,
in Modelling of Stellar Atmospheres, ed. N.E. Piskunov, W.W. Weiss, D.F. Gray, IAU Symp. 210, F3
\bibitem [2004] {NAHGH04}
Nowotny, W., Aringer, B., H\"ofner, S., et al. 2004,
in Proc. of ESO Workshop "High-resolution IR spectroscopy in Astronomy", ed. H.U. K\"aufl, R. Siebenmorgen, A. Moorwood, ESO Astrophysics Symposia, Springer, p.291
\bibitem [2005] {NowLH05}
Nowotny, W., Lebzelter, T., Hron, J., \& H\"ofner, S. 2005, A\&A accepted (Paper \,II)
\bibitem [1969] {PhilF69}
Phillips, J.G., \& Freedman, R.S. 1969, PASP, 81, 521
\bibitem [2001] {RichW01}
Richter, He., \& Wood, P.R 2001, A\&A, 369, 1027
\bibitem [2003] {RWWBS03}
Richter, He., Wood, P.R, Woitke, P., Bolick, U., \& Sedlmayr, E. 2003, 
A\&A, 400, 319
\bibitem [1984] {RiCHJ84}
Ridgway, S.T., Carbon, D.F., Hall, D.N.B., \& Jewell, J. 1984, ApJS, 54, 177
\bibitem [1991] {RoulM91}
Rouleau, F., \& Martin, P.G. 1991, ApJ, 377, 526
\bibitem [1950] {Sanfo50}
Sanford, R.F. 1950, ApJ, 111, 270
\bibitem [2003] {SchWS03}
Schirrmacher, V., Woitke, P., \& Sedlmayr, E. 2003, A\&A, 404, 267
\bibitem [1992] {Schol92}
Scholz, M., 1992, A\&A, 253, 203
\bibitem [2000] {SchoW00}
Scholz, M., \& Wood, P.R. 2000, A\&A, 362, 1065
\bibitem [1994] {Sedlm94}
Sedlmayr, E. 1994,
in Molecules in the Stellar Environment, IAU Coll. 146, ed. U.G. J{\o}rgensen, Lecture Notes in Physics 428, Springer, p.163
\bibitem [2004] {SimiW04}
Simis, Y., \& Woitke, P.
2004, in Asymptotic Giant Branch Stars, ed. H.J. Habing, H. Olofsson, Springer, ch.6, p.291
\bibitem [1988] {Tsuji88}
Tsuji, T. 1988, A\&A, 197, 185
\bibitem [1996] {WallH96}
Wallace, L., \& Hinkle, K. 1996, ApJS, 107, 312
\bibitem [1975] {Walle75}
Wallerstein, G. 1975, ApJS, 29, 375
\bibitem [1985] {Walle85}
Wallerstein, G. 1985, PASP, 97, 994
\bibitem [1998] {Winds98}
Windsteig, W. 1998, Master thesis, Vienna University of Technology, Austria
\bibitem [1998a] {WiHAD98}
Windsteig, W., H\"ofner, S., Aringer, B., \& Dorfi, E.A. 1998a,
in Proc. of ESO Workshop "Cyclical Variability in Stellar Winds", ed. L. Kaper, A.W. Fullerton, ESO Astrophysics Symposia, Springer, p. 308  
\bibitem [1998b] {WiLHD98}
Windsteig, W., Loidl, R., H\"ofner, S., \& Dorfi, E.A. 1998b,
in Fundamental Stellar Properties: The interaction between observation and theory, ed. T.R. Bedding, IAU Symp. 189, p. 172
\bibitem [1999] {WiALH99}
Windsteig, W., Aringer, B., Lebzelter, T., \& H\"ofner, S. 1999,
in Asymptotic Giant Branch Stars, ed. T. Le Bertre, A. Lebre, C. Waelkens, IAU Symp. 191, p. 614
\bibitem [2000] {WiKGS00}
Winters, J.M., Keady, J.J., Gauger, A., \& Sada, P.V. 2000, A\&A, 359, 651
\bibitem [1998] {Woitk98}
Woitke, P. 1998, in Cyclical Variability in Stellar Winds, ed. L. Kaper, A.W. Fullerton, ESO Astrophysics Symposia, Springer, p.278
\bibitem [2003] {Woitk03}
Woitke, P. 2003, in Modelling of Stellar Atmospheres, ed. N.E. Piskunov, W.W. Weiss, D.F. Gray, IAU Symp. 210, p.387
\bibitem [1988] {Yorke88}
Yorke, H.W. 1988, in Radiation in moving gaseous media, ed. Y. Chmielewski, T. Lanz, 18$^{th}$ Advanced Course of the Swiss Society of Astrophysics and Astronomy (Saas-Fee Course), p.193

\end{thebibliography}
\end{document}